\documentclass[aoas,MSNbibl,nameyear,dvips]{arximspdf}
\usepackage{graphicx}

\doi{10.1214/11-AOAS471}
\volume{5}
\issue{3}
\pubyear{2011}
\firstpage{2150}
\lastpage{2168}

\makeatletter
\def\@bmisc[#1]{%
  \get@battribute{unstr}%
  \common@pub@types%
  \let\bauthor\bbl@bauthor%
  \let\bhowpublished\@firstofone%
  \def\borganization##1{{\bauthor@style ##1}}%
}

\newcommand{\n}[1]{\bolds{#1}}
\newcommand{\cal}{\mathcal}
\makeatother

\begin{document}
\begin{frontmatter}

\title{A Bayesian Joinpoint regression model with an unknown number of break-points\thanksref{T1,T2}}
\runtitle{Joinpoint regression with unknown break-points}

\begin{aug}
\author[A]{\fnms{Miguel A.} \snm{Martinez-Beneito}\corref{}\ead[label=e1]{martinez\_mig@gva.es}},
\author[B]{\fnms{Gonzalo}~\snm{Garc\'{\i}a-Donato}\ead[label=e2]{Gonzalo.GarciaDonato@uclm.es}}
\and
\author[C]{\fnms{Diego} \snm{Salmer\'{o}n}\ead[label=e3]{diego.salmeron@carm.es}}

\runauthor{M. A. Martinez-Beneito, G. Garc\'{\i}a-Donato and D. Salmer\'{o}n}

\affiliation{Centro Superior de Investigaci\'{o}n  en Salud P\'{u}blica,
Universidad de Castilla la Mancha and Centro Superior  de Investigaci\'{o}n en Salud P\'{u}blica,
and
Consejeria de Sanidad and CIBER de Epidemiolog\'{\i}a  y Salud P\'{u}blica (CIBERESP)}
\address[A]{M. A. Martinez-Beneito\\
Centro Superior de Investigaci\'{o}n\\ \quad en Salud P\'{u}blica\\
Valencia\\
Spain\\\printead{e1}} 
\address[B]{G. Garc\'{\i}a-Donato\\
Universidad de Castilla la Mancha\\
Albacete\\
Spain\\
and\\
Centro Superior\\ \quad de Investigaci\'{o}n en Salud P\'{u}blica\\
Valencia\\
Spain\\
\printead{e2}}
\address[C]{D. Salmer\'{o}n\\
Servicio de Epidemiolog\'{i}a\\
Consejeria de Sanidad\\
Murcia\\
Spain\\
and\\
CIBER de Epidemiolog\'{\i}a\\ \quad y Salud P\'{u}blica (CIBERESP)\\
Spain\\
\printead{e3}}
\end{aug}
\thankstext{T1}{Supported in part by Ministerio de Educaci\'{o}n y
Ciencia; Contract Grants MTM2007-61554, MTM2010-19528.}

\thankstext{T2}{Supported in part by Fondo de Investigaciones
Sanitarias. Instituto de Salud Carlos III; Contract Grant ISCIII06-PI1742.}

\received{\smonth{11} \syear{2009}}
\revised{\smonth{3} \syear{2011}}

%
\begin{abstract}
Joinpoint regression is used to determine the number of segments needed
to adequately explain the relationship between two variables. This
methodology can be widely applied to real problems, but we focus on
epidemiological data, the main goal being to uncover changes in the
mortality time trend of a specific disease under study. Traditionally,
Joinpoint regression problems have paid little or no attention to the
quantification of uncertainty in the estimation of the number of
change-points. In this context, we found a satisfactory way to handle
the problem in the Bayesian methodology. Nevertheless, this novel
approach involves significant difficulties (both theoretical and
practical) since it implicitly entails a model selection (or testing)
problem. In this study we face these challenges through (i) a novel
reparameterization of the model, (ii) a conscientious definition of the
prior distributions used and (iii)~an encompassing approach which
allows the use of MCMC simulation-based techniques to derive the
results. The resulting methodology is flexible enough to make it
possible to consider mortality counts (for epidemiological
applications) as Poisson variables. The methodology is applied to the
study of annual breast cancer mortality during the period 1980--2007 in
Castell\'{o}n, a~province in Spain.
\end{abstract}

%
\begin{keyword}
\kwd{Bayesian statistics}
\kwd{model selection}
\kwd{Bayes factors}
\kwd{Joinpoint regression}
\kwd{epidemiological time series}.
\end{keyword}

\end{frontmatter}

\clearpage
\section{Introduction}\label{sec1}
Joinpoint regression is a statistical modeling technique that explains
the relationship between two variables by means of a segmented linear
regression constrained to be continuous everywhere, in particular, in
those places where the slope of the regression function changes. This
technique is widely applied to the modeling of time trends in mortality
or incidence series in epidemiological studies. In these applications
the number (if any) and the location of the changes in trends (known as
change-points or joinpoints) is usually unknown, the main goal being to
assess their existence and determine their location. Joinpoint
regression is applied in a wide variety of contexts. Nevertheless,\ for
clarity of presentation of the main ideas in this paper, the
above-mentioned epidemiological setting is assumed throughout the paper.

The underlying problem in this context is a model selection (or
testing) problem, with uncertainty about which is the model (number of
change-points) that (most likely) produced the data observed. In this
paper we approach the problem from a Bayesian perspective which is
fully detailed below. 
Nevertheless, the predominant techniques in the context of Joinpoint
regression are frequentist. Within them, the main goal is to find ``the
smallest number of joinpoints such that, if one more joinpoint is
added, the improvement is not statistically significant'' [\citet{JoinpointProgram}]. This goal is usually met by means of nonparametric
permutation tests [\citet{KimFayea2000}], the final result of that
analysis being the determination of the model that meets the former
condition and the estimation of its parameters and their variability.
The National Cancer Institute (NCI) of the United States has developed
a~tool to carry out these kinds of analyses [\citet{JoinpointProgram}].
This software has become a standard tool in epidemiological literature;
see, for example, \citet{CayuelaRodriguez-Dominguezea2004},
 \citet{BosettiLeviea2005},
 \citet{StracciCanosaea2007},
 \citet{Karim-KosVriesea2008},
 \citet{BosettiBertuccioea2008},
 \citet{QiuKatanodaea2009}.

The proposed permutation test approach has, in our opinion, two main
limitations:

\begin{longlist}[(ii)]
\item[(i)] The underlying model selection criterion selects the
simplest model such that if a new joinpoint is added, it does not yield
a statistically significant improvement. Therefore, the model selected
is not the most likely one, but it is chosen according to both its
capacity to describe the time trend within the data and the
informativeness of data to highlight temporal changes (in the case of
mortality or disease incidence time series, this last feature will
depend heavily on the average number of annual observed cases in the
data set). Hence, in a situation when the data is not very informative,
the criteria is clearly biased (by definition) toward more simplistic
models. This conservative behavior of the permutation procedure could
seem reasonable to some authors [\citet{KimFeuer2009}], but we find
this systematic tendency unsatisfactory. Instead, in this situation we
would expect to be ``informed'' that a number of alternative models are
equally plausible.
\item[(ii)]
It is hard to quantify to what extent one selected model is more likely
than others. In practice, a single model is chosen, with a fixed number
of joinpoints, regardless of whether the choice is much more likely or
not than the other alternatives. As a consequence, the main goal of the
inference in these kinds of models (how many joinpoints can adequately
explain the time trend that we are observing?) lacks an estimate of its
variability in contrast to the remaining parameters of the selected
model. This is because the asymptotics of the number of joinpoints is
an involved issue. Relevant advances in this regard include the work of
\citet{Yao1988},
 \citet{LiuZidek1997},
 \citet{KimFeuer2009}, focusing mainly on
conditions where estimators of this parameter are consistent (converge
to the parameter as the sample size increases).
\end{longlist}

There is some previous work devoted to the application of Bayesian
ideas to Joinpoint problems. \citet{CarlinGelfandea1992} is one of the
pioneering contributions in this area, proposing the application of
MCMC methods to fit these kinds of models. \citet{MorenoTorresea2005}
derived the intrinsic priors for a possibly heteroscedastic normal
model under the assumption of a fixed change-point. In a similar
problem but under homoscedasticity, more recently, \citet{BayarriGarcia-Donato2007} explicitly derived the Zellner--Siow priors
[\citet{ZellSiow80}; \citet{Zellner1984}]. Bearing in mind its epidemiological
application, \citet{TiwariCroninea2005} describe the calculus of Bayes
factors for model selection in Gaussian joinpoint models. A common
limitation of the application of all these studies to epidemiological
time series modeling is the assumption of normal errors. This
restriction is relaxed in the work of \citet{GhoshBasuea2009}, who
propose semiparametric regression models by means of Dirichlet
processes. Nevertheless, the original data in mortality studies are
usually the annual observed death counts, for which in this paper we
assume a Poisson regression model to take into account the discrete
nature of these data. As far as we know, the only Bayesian model
selection approach that considers Poisson counts in the context of
Joinpoint regression is \citet{GhoshHuangea2009}. These authors also
acknowledge the advantage of the Poisson assumption especially when the
observed counts are smaller, in which case the Gaussian modeling of
incidence or mortality rates is clearly not convenient. That is
probably the work most closely related to ours. Nevertheless, our
approach differs from it in at least two main ways. First, \citet{GhoshHuangea2009} model the hazard rate in a context of relative
survival, while we propose a model for the, more usual in this context,
incidence or death rates (see Section~\ref{sec3.1}). Second, the model selection
performed in \citet{GhoshHuangea2009} is based on popular model
selection criteria like CPO and DIC, but not on posterior probabilities
as we do. The advantages of posterior probabilities over other criteria
are a straightforward interpretation and the richness of the results
produced (e.g., to produce predictions). These are put in practice in
Section~\ref{sec4}.\looseness=-1

The Bayesian approach is straightforward at least conceptually. Bayes
factors allow the selection of models from among several alternatives,
strictly according to their posterior probabilities. Furthermore,
through Bayes factors it is possible not only to select one of the
models entertained, according to the evidence (posterior probability)
provided by the data, but also to quantify the difference between the
one selected and the remaining competing models. In a broad sense,
Bayes factors make it possible to evaluate the uncertainty involved
with the selection made. Furthermore, their use is the basis for what
has been called Model Averaging [\citet{Clyde1999}], under which it is
possible to average the fit of all the models weighted by their
posterior probabilities. Therefore, uncertainty in the selection of the
``correct'' model is propagated to the inferential exercise.

The goal of this study is to propose a Joinpoint regression modeling
that evaluates and incorporates the uncertainty in both model selection
and model parameters into the analysis. We found the Bayesian approach,
for the aforementioned reasons, a very appealing way of doing so. Of
course, the Bayesian approach to model selection problems is not free
of difficulties, as nicely explained in \citet{BergerPericchi2001}.
These can basically be summarized in two main problems: a~strong
influence of the prior on the results and a very challenging numerical
problem to compute these results. Much of the material presented in
this paper focuses on how we manage to overcome these problems.

The rest of the paper is organized as follows: Section~\ref{sec2} introduces a
reparameterization of the usual Joinpoint regression model which will
be really convenient as a first step to assign prior distributions. In
Section~\ref{sec3}, starting from the reparameterization proposed in Section~\ref{sec2},
we introduce a Joinpoint modeling proposal with an unknown number of
change-points. That proposal will be carried out as a variable
selection process [\citet{GeorgeMcCulloch1993}; \citeauthor{DellaportasForsterea2000}
(\citeyear{DellaportasForsterea2000,DellaportasForsterea2002})], and prior
distributions will be discussed in detail in order to get reasonable
results from the previous model selection problem. 
In Section~\ref{sec4} our new model will be applied to the study of breast
cancer mortality in the Spanish province of Castell\'{o}n to illustrate
the possibilities of the Bayesian approach for exploiting the results
from the inference. Finally, in Section~\ref{sec5} we will summarize the main
advances of our model and some future lines of work will be outlined.

\section{A convenient parameterization of the joinpoints}\label{sec2}
Suppose that we want to describe the behavior of the variable $\{Y_i;
i=1,\ldots,n\}$ as a function of the explanatory covariate $\{t_i,
i=1,\ldots,n\}$. We find it convenient to assume that $t_i$ represents
time, although it could represent any other magnitude. It is usual to
model the presence of $J$ change-points (locations in which the
functional describing the relationship between $Y$ and $t$ changes)
through the expected value of the dependent variable as
%
\begin{equation}\label{JoinClasic}
g(E(Y_i|t_i))=\alpha+\beta_0\cdot t_i+ \sum_{j=1}^J\beta_j\cdot(t_i-\tau_j)^+
\end{equation}
for certain $g$, a linking function. In this equation, $\tau_j$
represents the location of the $j$th change-point and $(\theta)^+$ is
$\theta$ if $\theta>0$, and 0 otherwise. We label the model in (\ref{JoinClasic}) as $M_J$, to make it explicit that it contains exactly
$J$ change-points. Similarly, we call the model with no change-points $M_0$.

In what follows, we assume that the maximum number of change-points is~$J^*$,
a number which is fixed (more on this aspect later). Then the
problem that we face is to find the posterior probability of each of
the models in $\{M_0,M_1,\ldots,M_{J^*}\}$ and in the case that a single
model needs to be selected, the one with the highest posterior
probability is preferred.

In the Bayesian framework, the distinction between common and new
parameters for the assignment of prior distributions [first used by
\citet{Jeffreys1961}] is crucial. The terminology is very intuitive:
common parameters appear in all the competing models while the new
parameters are model-specific. In the problem above, $\alpha$ and $\beta
_0$ are common and the remaining $\beta$'s are new parameters.

As we fully describe in the following section, the scheme that we adopt
for the assignment of the priors proposes the same marginal
(noninformative) prior for the common parameters. Clearly, using the
same prior (either subjectively elicited or not) for common parameters
does not make sense unless the meaning of these parameters does not
change throughout the different models entertained. Unfortunately,
common parameters do not (in general) represent the same magnitude,
this being a major difficulty in Bayesian model selection [see \citet{BergerPericchi2001}]. Although it is hard to establish precise
conditions under which common parameters have the same meaning (a
notion which can be viewed as subjective), it is clear that, as it
stands, this is not the case for the problem presented above. For
instance, in $M_0$, $\alpha$ and $\beta_0$ represent the parameters
explaining the global trend of the series. Nonetheless, in $M_1$ (a
model with one joinpoint) $\alpha$ and $\beta_0$ are the intercept and
the slope respectively of a line adjusted \textit{up to} the change-point
($\tau_1$), $\beta_0+\beta_1$ being the slope from the joinpoint on. We
propose an alternative parameterization, for which we argue that the
hypothesis of common parameters with a same meaning across models holds
reasonably.
More concisely, let
%
\begin{equation}\label{JoinNuevo}
E(Y_i|t_i)=\alpha+\beta_0\cdot(t_i-\bar{t})+ \sum_{j=1}^J\beta_j
\mathcal{B}_{\tau_j}(t_i),
\end{equation}
where $\mathcal{B}_{\tau_j}(t)$, which we call the \textit{break-point}
or the \textit{break-point centered at
$\tau_j$}, is defined as the following piecewise linear function:
\[
\mathcal{B}_{\tau_j}(t)=
\cases{\displaystyle
a_{0j}+b_{0j} \cdot t,& \quad  $\forall t\le\tau_j $,\cr\displaystyle
a_{1j}+b_{1j} \cdot t  ,& \quad  $\forall t>\tau_j$,
}
\]
restricted to a number of conditions which are fully specified below.
What we finally obtain is, conditioned on the joinpoints, a known
function of the original regressors
at times $\{t_1,t_2,\ldots,t_n\}$ and describing a peak at the moment
$\tau_j$. Alternatively, the set of break-points can be interpreted as
a base of continuous piecewise linear functions generating the
joinpoints needed to describe the time trend in the data. The linear
component in (\ref{JoinNuevo}) has been introduced as $(t_i-\bar{t})$
(instead of $t_i$) to avoid the dependency between $\alpha$
and~$\beta_0$.\looseness=-1

The conditions imposed on $\mathcal{B}_{\tau_j}(t)$ are as follows:
\begin{itemize}
\item$\lim_{t \to\tau_j^-}\mathcal{B}_{\tau_j}(t)=\lim_{t \to\tau
_j^+}\mathcal{B}_{\tau_j}(t)$,
that is, $\mathcal{B}_{\tau_j}(t)$ is a continuous function. Hence, it
is guaranteed that the regression function (\ref{JoinNuevo}) is
continuous all around, independently of the number of break-points.

\item$\sum_{i=1}^n \mathcal{B}_{\tau_j}(t_i)=0$, that is, the sum of
the elements of the break-point evaluated in all the points observed
must be zero. This way the addition of any break-point in the model
would not alter the mean value of the regression function and it would
not change the meaning and the estimation of parameter $\alpha$ across
models. In other words, this condition can be understood as if the new
break-points are imposed to be geometrically orthogonal to the
intercept term.

\item$\sum_{i=1}^n \mathcal{B}_{\tau_j}(t_i)\cdot t_i=0$, that is, the
slope of the break-points along the whole period of study is zero. This
way the addition of any break-points in the model would neither alter
the slope of the regression function nor would it change the meaning
and the estimate of parameters $\beta_0$ across models. In other words,
this condition can be understood as if the new break-points are imposed
to be geometrically orthogonal to the slope term.

\item$\mathcal{B}_{\tau_j}(\tau_j)=1$, so that the parameter $\beta_j$
has the role of measuring the magnitude
of the break-point in the location where the change in the tendency
takes place. Hence, $\beta_j$ has to be
interpreted as the value of the deviation produced at $\tau_j$ as a
consequence of including this
break-point in the model. Without a restriction of this kind the value
of $\beta_j$ would not be identifiable.
\end{itemize}

Subject to these conditions and given $\tau_j$, the function $\mathcal
{B}_{\tau_j}(t)$ is
unambiguously determined (details are provided as supplementary
material [\citet{Martinez-BeneitoGarcia-Donatoea2011b}]).

The main motivation for introducing the above basis of functions is to
improve robustness over prior specification by reparameterization as is
further explained below. In this sense, our approach is different from
the other basis of functions used [like MARS in \citet{Friedman1991} and
\citet{NottKukea2005}] where the aim was to approximate surfaces which
could potentially be highly nonlinear on their multiple arguments. With
this parameterization, it is now reasonable to assume that the common
parameters have the same meaning: in all models, $\alpha$ and $\beta_0$
are parameters of a line representing the global trend. The remaining
parameters are used to modify this common line to incorporate changes
in the trend without changing the original meaning of $\alpha$ and
$\beta_0$ in $M_0$.

\begin{figure}

\includegraphics{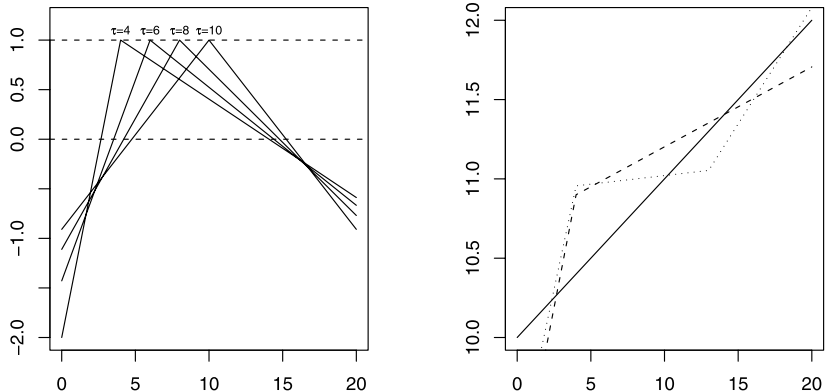}

\caption{Left: Shape of different break-points as a function
of their change-point locations. Right: Trends with 0 (solid
line), 1 (dashed line) and 2 (dotted line) change-points. All these
functions have the same mean value and slope for the whole period.}
\label{figuraBreaks}
\label{fig1}\end{figure}

As an illustration, a graphical representation of several break-points,
corresponding to different locations of the change-points (for 20
observed values), is shown in Figure \ref{figuraBreaks} (left). As can
clearly be seen, all break-points are equal to 1 at their corresponding
change-points, are zero on average and their slope is also null.
Following this same context, Figure \ref{figuraBreaks} (right) shows
how the inclusion of break-points modifies a straight line common to
all models ($\alpha+\beta_0\cdot t$). In this figure we have plotted
the line $y(t)=10+0.1\cdot t$, jointly with the functions
$y(t)=10+0.1\cdot t+0.5\cdot\mathcal{B}_5(t)$ and $y(t)=10+0.1\cdot
t+0.5\cdot\mathcal{B}_5(t)-0.3\cdot\mathcal{B}_{13}(t)$. The effect of
including break-points over a common regression function is the torsion
of this function, instead of a divergence from a certain place as was
done in the parameterization in (\ref{JoinClasic}). In other words, the
effect of the introduced break-points is a~global modification of the
regression curve, as opposed to the previous parameterization which
produced a local modification, changing the meaning of the parameters
$\alpha$ and $\beta_0$ depending on whether the break-points are (or
are not) included in the model and how many of them are included.
%

Additionally, a number of nice side effects are derived from this
parameterization:
\begin{itemize}
\item With the definition of the break-points, the columns of the
design matrix corresponding to the common parameters are orthogonal to
the columns of the new parameters. Hence, we can reasonably expect the
Fisher information to be approximately block diagonal (of course, it
depends on the particular distribution assumed for $Y$, and will be
true for the normal case). In this scenario, it is known that the Bayes
factors (and consequently any other result derived from them, like the posterior
probabilities) are quite robust to the prior distribution used for
common parameters [see \citet{Jeffreys1961}; \citet{KassVaidyanathan1992}]. This has been used as a justification for the
strategy of using noninformative priors, possibly improper, for common
parameters [see \citet{LiangPauloea2008};
 \citet{BayarriGarcia-Donato2008}].
We use a similar approach, as we explain in the next section.

\item With this reparameterization, the common parameters $\alpha$ and
$\beta_0$ conserve their meaning regardless of the model (i.e.,
regardless of the number of joinpoints). This allows us to make
inferences about them in a model-averaged way, that is, taking into
account the uncertainty regarding which is the true model. We put this
into practice in Section~\ref{sec4} where we draw inferences on these parameters.
\end{itemize}

\section{A Joinpoint regression model with an unknown number of change-points}\label{sec3}
\subsection{An encompassing model}\label{sec3.1}
To address the question of how many joinpoints are needed, or,
equivalently, to select from among
$\{M_0,M_1,\ldots,M_{J^*}\}$, we introduce an encompassing model in
which all these models are
nested. We also assume the scenario of modeling epidemiological series
of mortality or incidence
counts of a certain disease, since this is the most extended
application of Joinpoint models.
Nevertheless, the main ideas that we present hold for other types of
data, not necessarily
counting data.

Let $Y_i$ be the number of cases of a specific disease observed
during a~period of time, represented by $t_i$. To account for the
discrete nature of these values, we suppose that
\[
Y_i\sim \operatorname{Poisson}(\mu_i), \qquad    i=1,\ldots,n,
\]
whose mean is defined as
%
\begin{equation}\label{LogMediaModelo}\log(\mu_i)=\log(P_i)+\alpha+\beta
_0\cdot(t_i-\bar{t})+\sum_{j=1}^{J^*} \delta_j\cdot\bigl(\beta_j \cdot
\mathcal{B}_{\tau_j}(t_i)\bigr),
\end{equation}
where $P_i$ is the population under study during year $i$, $\{\delta_j;
j=1,\ldots,J^*\}$ are binary variables which include (for $\delta_j=1$) or
exclude (for $\delta_j=0$) each change-point in the model and $\tau_j$
is the unknown location of the $j$th break-point.
This model can be interpreted as an encompassing model, which contains
all possible joinpoint models unambiguously identified through $\n\delta
=(\delta_1,\ldots,\delta_{J^*})$,\vspace*{1pt} to explain the data. In particular,
$M_0\equiv\{\n\delta=\n0\}$, $M_1\equiv\{\n\delta\dvtx \allowbreak\sum
_{j=1}^{J^*}\delta_j=1\},\ldots$ and $M_{J^*}\equiv\{\n\delta=\n1\}$.\vspace*{1pt}
Hence, the posterior probability over the model space $\{
M_0,M_1,\ldots,M_{J^*}\}$ is completely specified through the posterior
distribution of $\n\delta$. A detailed description of the computation
of $\pi(\n\delta\vert\mathbf y)$ is provided as supplementary material
[\citet{Martinez-BeneitoGarcia-Donatoea2011b}]. The proposal in (\ref{LogMediaModelo})
can in a way be seen as an \mbox{order-2} regression spline
[\citet{HastieTibshiraniea2009}], as this is a~piecewise linear
continuous regression function.

To avoid identifiability problems, we impose a number of restrictions
on the locations of the change-points on which there is broad consensus
in the related literature. These are imposed so as to ensure a minimum
distance between change-points and a restriction of order. In
particular, we assume that the parametric space for $\n\tau=(\tau_1,\tau
_2,\ldots,\tau_{J^*})$, which we call $\Omega$, is
\[
\Omega=\{(\tau_1,\ldots,\tau_{J^*})\dvtx   t_1+d<\tau_1, \tau_1+d<\tau_2,
\tau_2+d<\tau_3,\ldots,\tau_{J^*}+d<t_n \}.
\]

With this assumption, there is a minimum number of periods $d$ between
any two change-points. By default, we use $d=2$ in our applications.
Notice that this restriction avoids the existence of two or more
change-points between consecutive observations.
Other ways of implementing this restriction have been proposed, as, for
example, in \citet{GhoshBasuea2009}.
Notice that $\Omega$ is a bounded set in ${\cal R}^{J^*}$.

\subsection{The prior distribution}\label{sec3.2}
Let $\pi(\alpha,\beta_0,\n\delta,\n\tau,\n\beta)$ be the prior
distribution of the parameters in
the model (\ref{LogMediaModelo}). We express this prior as
\[
\pi(\alpha,\beta_0,\n\beta,\n\tau,\n\delta)=
\pi(\n\beta\vert
\alpha,\beta_0,\n\delta,\n\tau)\pi(\alpha,\beta_0)\pi(\n\tau)\pi(\n
\delta),
\]
where $\n\beta=(\beta_1,\ldots,\beta_{J^*})$.

The prior $\pi(\alpha,\beta_0)$ corresponds to common (with the same
meaning, as argued in the previous section) parameters in all models.
Under this condition,
it is common to use a noninformative prior [see, e.g.,
\citet{BergerPericchi2001};
 \citeauthor{BayarriGarcia-Donato2007} (\citeyear{BayarriGarcia-Donato2007,BayarriGarcia-Donato2008})], in this case the constant prior: $\pi
(\alpha,\beta_0)\propto1.$

All the other parameters are not common and it is well known [\citet{BergerPericchi2001}] that the posterior distribution is very
sensitive to their prior distribution. In particular, noninformative
(improper) or vague priors would produce arbitrary Bayes factors.

Our approach to assigning $\pi(\n\beta\vert
\alpha,\beta_0,\n\delta,\n\tau)$ is based on an approximation of the
divergence based (DB)
prior, introduced by \citet{BayarriGarcia-Donato2008} as a broad
generalization of the pioneering
ideas of \citet{Jeffreys1961},
 \citet{ZellSiow80} and \citet{Zellner1984}. When
comparing two nested models, an approximation of the DB prior for the
new parameters in the complex model (possibly reparameterized to range
within the real line) is a
heavy tailed density, centered at zero and scaled by the inverse of the
corresponding block of
the unitary Fisher information matrix evaluated in the simpler model.
This would lead us to the proposal
%
\begin{equation}\label{Cauchy}
\pi(\n\beta\vert\alpha,\beta_0,\n\delta,\n\tau)=\mbox{Cauchy}_{J^*}(\n
\beta\vert\n0,\n\Sigma),
\end{equation}
the matrix $\n\Sigma$ to be specified. Using a hierarchical scheme,
this is equivalent to the proposal
\[
\pi(\n\beta\vert\alpha,\beta_0,\n\delta,\n\tau,\gamma)=\mbox
{Normal}_{J^*}(\n\beta\vert\n0,\gamma\n\Sigma), \qquad  \gamma\sim
\operatorname{Gamma}^{-1}(0.5,0.5),
\]
where $\gamma$ acts as a ``mixing'' parameter.

In our problem, it is easy to see that the block (corresponding to $\n
\beta$) of the Fisher information matrix of the encompassing model (\ref{LogMediaModelo}) evaluated at $\n\beta=\n0$ is ${\cal I}=\n\Delta
\mathbf
B^t\mathbf W\mathbf B\n\Delta$, where $\mathbf B=\{B_{\tau_j}(t_i)\}$ (the matrix of
covariates),
$\mathbf W=\operatorname{diag}\{w_i\}$ with $w_i=P_i \operatorname{exp}(\alpha+\beta
_i(t_i-\bar{t}))$ and $\n\Delta=\operatorname{diag}(\n\delta)$. Clearly,
${\cal I}$~is not (for every $\n\delta$) a positive-definite matrix, so it
cannot be used directly to define $\n\Sigma$ above. Instead, we propose
%
\begin{equation}\label{Sigma}
\n\Sigma=n \bigl(\n\Delta\mathbf B^t\mathbf W\mathbf B\n\Delta+\operatorname{diag}(\mathbf B^t\mathbf W\mathbf
B-\n\Delta\mathbf B^t\mathbf W\mathbf B\n\Delta) \bigr)^{-1},
\end{equation}
which is (for every $\n\delta$) a positive-definite matrix.

As we argue below, the resulting proposed prior (\ref{Cauchy}) with a
scale matrix as in (\ref{Sigma}) has a very interesting interpretation.
Given a particular $\n\delta^*$, with, say, $\sum\delta_i^*=J$, let $\n
\beta_\delta$ be the $J$-dimensional subvector of $\n\beta$ which
corresponds to the nonnull $\delta_i$'s. Similarly, denote $\n\beta
_{\delta^c}$ as the remaining parameters in $\n\beta$. Finally, $\mathbf
B_\delta$ denotes the matrix with the columns in $\mathbf B$ which
corresponds to the nonnull $\delta_i$'s. Hence, it can easily be seen
that $\n\beta_\delta$ and $\n\beta_{\delta^c}$ are independent,
conditional on the mixing parameter $\gamma$. In
fact,
\[
\pi(\n\beta\vert\alpha,\beta_0,\n\delta=\n\delta^*,\n\tau,\gamma)=\mbox
{Normal}_{J}(\n\beta_\delta\vert\n0,\gamma n  (\mathbf B_\delta^t\mathbf W\mathbf
B_\delta )^{-1})
f(\n\beta_{\delta^c}\vert\alpha,\beta_0,\n\tau,\gamma),
\]
where
\[
f(\n\beta_{\delta^c}\vert\alpha,\beta_0,\n\tau,\gamma)= \prod_{\{
i\dvtx \delta_i=0\}} \mbox{Normal}(\beta_i\vert0,\gamma n [\mathbf B^t\mathbf W\mathbf
B]_{ii}^{-1}).
\]

For every $\n\delta$, the joint prior $\pi(\n\beta\vert\alpha,\beta_0,\n
\delta,\n\tau)$ can be seen as the product of the (approximated) DB
prior for the \textit{active} parameters in the model,~times a~proper
density for those inactive parameters. This proper density has no
effect on the corresponding Bayes factors, thereby acting as a
pseudoprior [see, e.g., \citet{CarlinChib1995};
 \citet{HanCarlin2001};
 \citet{DellaportasForsterea2002}]. Nevertheless, as noticed previously in
the literature, these may have a great impact on the numerical results.
In our experience, partially described in the supplementary material
[\citet{Martinez-BeneitoGarcia-Donatoea2011b}], this particular form of
the pseudoprior leads to quite satisfactory results.

With this approach, the priors are in some sense defined
encompassingly. That is, instead of using one (DB) prior for each
possible submodel nested in (\ref{LogMediaModelo}), implicitly defined
by each $\n\delta$, we use a single prior which contains all these
priors. The main advantage of the resulting procedure is that it makes
it possible (as we describe in the next section) to use standard
estimation procedures (like MCMC) to easily solve the model selection
problem, whose direct resolution usually requires the help of
sophisticated numerical techniques.

The proposal for $\pi(\n\tau)$ is not as delicate an issue as the prior
for $\n\beta$. This is because the corresponding parametric space
$\Omega$ is a bounded set which ensures that $\pi(\n\delta)$ is proper
under very mild conditions. Hence, the normalizing constant is
unambiguously defined. 
In this situation, the constant prior is clearly a reasonable default choice:
$\pi(\n\tau)\propto1$ for $\n\tau\in\Omega.$

Finally, we introduce our proposal for $\pi(\n\delta)$, which is deeply
related to our prior beliefs over the model space $\{M_0,M_1,\ldots
,M_{J^*}\}$. One possibility would be to use independent Bernoulli
distributions for $\delta_i$ with a probability of success of 0.5.
Nevertheless, this would lead us to the $\operatorname{binomial}(J^*,0.5)$
distribution over the model space, clearly favoring those models with
around~$J^*/2$ joinpoints and also giving an undesirable important role
to the fixed value~$J^*$ (which is usually posed as arbitrarily large).

Instead, we experimented with two different proposals. In the first one
(which we refer to as \textit{Bayes1}), all models $M_i$ have the same
probability [i.e., $1/(J^*+1)$], this probability being equally
distributed over the same number of joinpoints. This prior distribution
can be formulated in a hierarchical way as
\[
\pi(\n\delta|\mathbf p)\propto\frac{\prod_{i=1}^{J^*} p_i^{\delta_i}
(1-p_i)^{(1-\delta_i)}}{  {
J^* \choose  \sum_{j=1}^{J^*} \delta_j
}
 }, \qquad
   \pi(p_i)=\operatorname{Beta}(1/2,1/2),  i=1,\ldots,J^*.
\]

The term in the denominator of the prior for $\n\delta$ yields the same
prior probability for any number of change-points in the model if
$p_1=\cdots =p_{J^*}=0.5$ (the prior expected values of these terms). This
hierarchical formulation, once the parameters $\mathbf p$ are integrated
out, can also be expressed as
\[
\pi_1(\n\delta)=(J^*+1)^{-1}{ \pmatrix{\displaystyle
J^* \vspace*{2pt}\cr\displaystyle  \sum_{j=1}^{J^*} \delta_j
}}^{-1}.
\]

The main drawback of this proposal is that,  a priori, the mean
number of joinpoints included in the model ($J^*/2$) depends on the
arbitrary quantity~$J^*$. While this is not a problem when $J^*$ is
carefully assigned, it could be an issue when this parameter is
arbitrarily assigned (as is the case in many studies). To avoid this
dependency on $J^*$ as much as possible, while keeping the essence of
the first proposal, we alternatively explore a slight modification on
$\pi_{1}(\n\delta)$. The prior (to be called \textit{Bayes2}) can be
defined hierarchically as
\begin{eqnarray*}
\pi(\delta_i|p_i)&=&\operatorname{Bernoulli}(p_i), \qquad
i=1,\ldots,J^*,
\\
\pi(p_i)&=&\operatorname{Beta}\bigl(1/2,(J^*-1)/2\bigr), \qquad   i=1,\ldots,J^*.
\end{eqnarray*}
With this second proposal, the prior expected number of
break-points\break
(1~break-point) does not depend on the value of $J^*$.

In the same way as for \textit{Bayes1}, this prior distribution can also
be expressed in a nonhierchical way, once the $p_i$ parameters are
marginalized, as
\[
\pi_2(\n\delta)=(J^*)^{-J^*}(J^*-1)^{J^*-\sum_{j=1}^{J^*}\delta_j}.
\]

Interestingly, as pointed out by an Associate Editor, for large $J^*$,
the prior probability of zero and one joinpoint tends to $e^{-1}\approx
0.37$, emphasizing the robustness of $\pi_2$ to the choice of $J^*$.

\section{Breast cancer mortality trend in Castell\'{o}n province}\label{sec4}
We analyze the breast cancer mortality time trend in Castell\'{o}n
province, for the period 1980--2007. Castell\'{o}n is one of the 50
provinces that make up Spain, where around 285,000 women resided in
2007 and with 62.4 women dying annually of breast cancer, on average,
during that period. Two facts have occurred in those years that have
presumably changed breast cancer mortality trends, which are the
progressive introduction of the Breast Cancer Screening Program in that
province in 1992 [\citet{VizcainoSalasea1998}], and the introduction
of new therapies for the treatment of this disease around the world, at
approximately the same time. In fact, there is some controversy over
which of these two factors could have a higher impact on mortality
variation [\citet{Peto1996}; \citet{BlanksMossea2000}; \citet{BerryCroninea2005}].
In any case, we would expect to find at least one joinpoint on the
breast cancer mortality trend studied. Moreover, changes in breast
cancer mortality trends have already been described for the first half
of the 1990s for other Spanish regions [\citet{AscunceMoreno-Iribasea2007};
 \citet{SalmeronCireraea2009}; \citet{PerezLacastaGregoriGomisea2010}].

\begin{figure}

\includegraphics{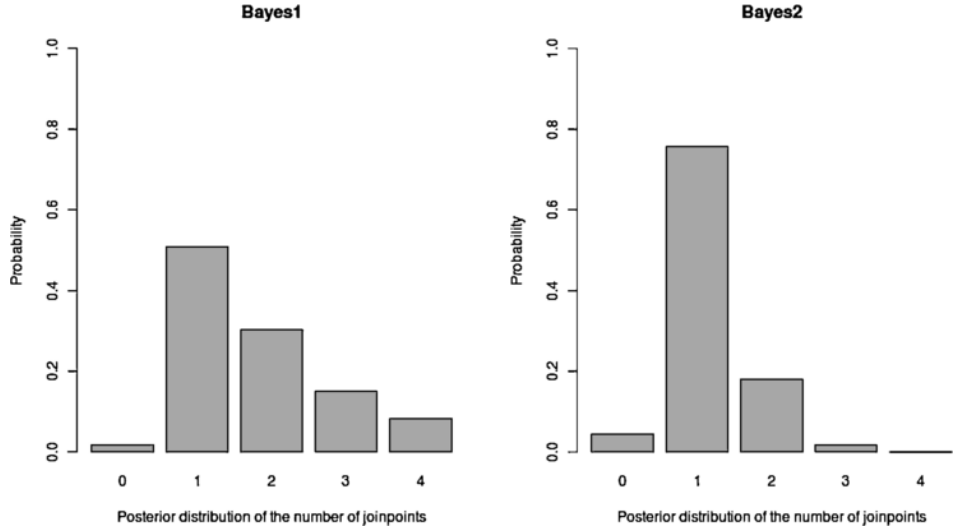}

\caption{Posterior distribution of the number of joinpoints in breast
cancer mortality trend in Castell\'{o}n Province for  {Bayes1} and
 {Bayes2} criteria.}
\label{fig2}
\end{figure}

Regarding Bayesian criteria, if we focus on the mode of the posterior
distribution of the number of joinpoints, in Figure~\ref{fig2} it can be seen
that both methods point to the existence of exactly one joinpoint,
indeed, \textit{Bayes1} yields a 51.0\% probability of one joinpoint,
while \textit{Bayes2} estimates that probability at 73.6\%. Moreover, the
probability of no joinpoint in the time trend for these two models is
1.9\% and 3.0\%, respectively. Hence, the probability of a ``simply
linear'' trend for breast cancer mortality is low.

%

On the other hand, we have used the \textit{Permutation} and \textit{BIC}
criteria, by means of the NCI tool. The \textit{Permutation} criterion
(the one suggested by the NCI tool's authors) does not find any
joinpoints in the period under study, possibly due to the limited
information on the trend provided by the low number of deaths by year.
The \textit{BIC} criterion, however, finds one joinpoint around 1995,
with a 95\% confidence interval: [1990, 2005]. As a~consequence, the
conservative behavior observed of the \textit{Permutation} criterion
makes it yield results which neither agree with the remaining criteria
nor with the results that we expected, based on further knowledge of
this cause of mortality. It has to be acknowledged that the Permutation
test yields a $p$-value of 0.0136 when testing 1 versus 0 joinpoints,
whereas the significance level of the test is 0.0125 (the result of the
Bonferroni correction of the original level 0.05), that is, the model
without any joinpoint is really close to being rejected. Therefore, the
arguments for selecting between the models with either none or just one
joinpoint are not conclusive at all for this example, but, on the
contrary, the consequences derived from that selection are dramatic.
This result warns against the danger of those criteria choosing
a~particular number of joinpoints and ignoring the uncertainty in that
choice. In our example the permutation test would presumably have led
to a wrong answer and from that moment on all the results from the
analysis would have been completely missleading, as the premises of the
chosen model are accepted as true and those from alternative models are
completely ignored.

%
\begin{figure}

\includegraphics{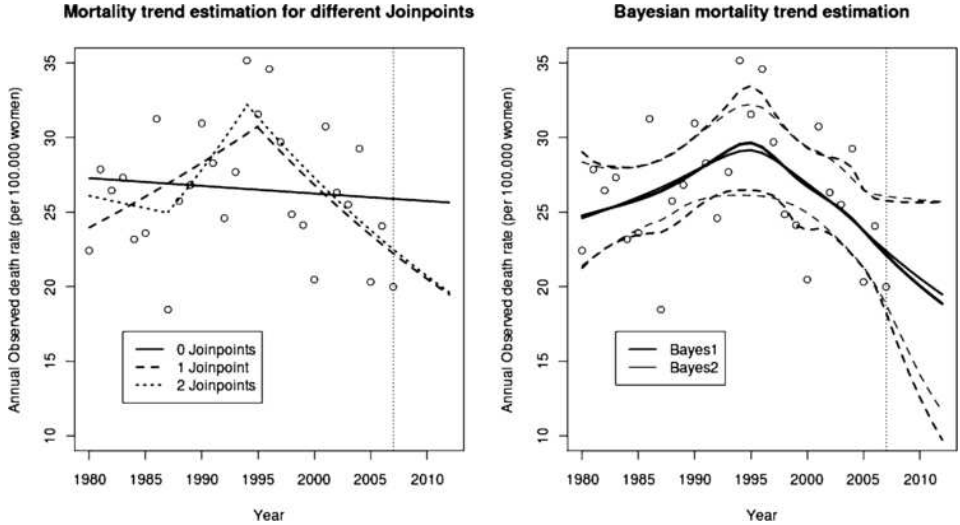}

\caption{Left: Least squares estimated trends for a different number of
joinpoints for the annual observed death rate (per 100,000 women).
Right: Estimated time trend for the annual observed death rate (per
100,000 women) for both  {Bayes1} and  {Bayes2}.}
\label{fig3}
\end{figure}

The left-hand side of Figure~\ref{fig3} shows the least squares estimated trend
for a different number of joinpoints (recall that the \textit
{Permutation} test chooses the curve with 0 joinpoints from among all
these). As can be noticed, fitted trends are quite different and the
repercussion of the choice of one or another curve, and as a
consequence of ignoring alternative models are dramatic. Those
consequences could be even worse in the case of temporal forecasting.
The former figure also shows the forecasted trend for every one of
these curves for the following 5 years (2008--2012) and, as can be
appreciated, predictions for the model without joinpoints diverge
completely from those of the models with joinpoints. As the Permutation
based criteria lacks an estimate of the probability of these scenarios,
they cannot be averaged and one of them has to be chosen with the
previously outlined risks. This is not the case of our proposed
methods. The right-hand side of Figure~\ref{fig3} shows the estimated trend by
both \textit{Bayes1} and \textit{Bayes2} where predictions based on
different numbers of joinpoints are averaged to provide a single answer
weighting all the scenarios considered. Moreover, as pointed out in
\citet{George1999}, the predictions derived with this Model Averaging
procedure are optimal in several senses, the square error loss being
one of them. Predicted trends for both Bayesian criteria are really
close, although the \textit{Bayes1} prediction depicts more detail than
the \textit{Bayes2} as a consequence of the higher number of joinpoints
that this model occasionally considers. From now on we will focus on
the results of \textit{Bayes1} to improve the clarity of the presentation.

%
\begin{figure}

\includegraphics{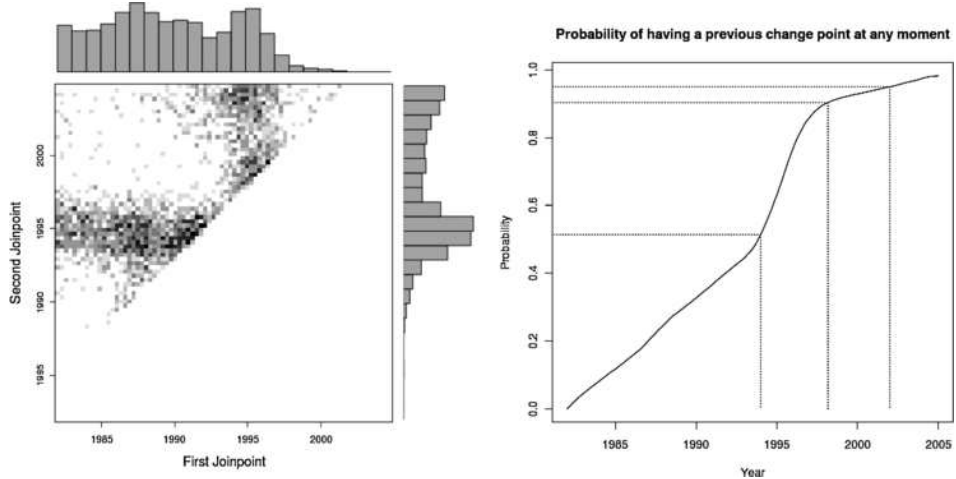}

\caption{Left: Joint posterior distribution of the location of
joinpoints conditioned to the model with just 2 joinpoints. Histograms
correspond to the marginal posterior distribution of the location of
the first and second joinpoint. Right: Probability of the trend having
described a change-point before any specific moment.}
\label{fig4}\end{figure}

The estimation of $\beta_0$ from \textit{Bayes1} yields a posterior mean
of $-$0.0017 and a 95\% probability interval $[-0.0077,0.0042]$.
Consequently, the global trend during the whole period studied has a
slope which is not very different from 0. The interest of this result
is that it is valid regardless of the true number of joinpoints
underlying the true trend, as we are not conditioning our results on a
specific number of change-points. Conversely, the frequentist results
condition all their conclusions on the selected model, and, therefore,
they are valid if and only if that model agrees with the reality.

Figure~\ref{fig4} (left) shows the joint posterior distribution of the location
of these two joinpoints. As can be noticed, this figure points out the
existence of one joinpoint from about 1993 to 1997 and another one less
precisely placed in the range from 1983 to 2005. For this second
joinpoint there are several places that seem to be likely to host the
second joinpoint. This information is much richer than that provided by
the least squares estimate in Figure~\ref{fig3} that just points to 1987 as the
most likely year for the second joinpoint. Focusing in the results in
Figure~\ref{fig4} (left), our impression is that reducing the results of the
2-joinpoints-fit to a single curve (although it corresponds to the
least squares fit) could be a really poor summary of all these results.
Moreover, the marginal distribution of the locations of both joinpoints
are multimodal and clearly asymmetric. Therefore, confidence intervals
obtained for these quantities under asymptotic approximations will have
to be treated with caution.

%

Finally, Figure~\ref{fig4} (right) shows, at every moment of the period under
study, the probability of the trend having described a change-point
before that precise moment. As it is evident that the curve converges
to 0.981, the complement of the probability of not describing a
joinpoint along the whole period of study. Moreover, from the start of
1994 the probability of already having described a joinpoint is higher
than 50\%, at the start of 1998 that probability is higher than 90\%
and we have to wait until 2002 to be sure, with a 95\% probability, of
having described at least one joinpoint. This way we can also measure
(without any asymptotic assumptions) the degree of certainty that we
have about the trend having described a change at any moment. It is
possible to derive many posterior summaries of this kind with this
approach. These will be able to answer most of the epidemiological
questions that could be of particular interest in real studies.

\section{Conclusions}\label{sec5}

This study has introduced a novel approach to Joinpoint analysis,
taking advantage of considering the model selection problem as an
inference problem on an encompassing model which contains all the
candidate alternatives. Moreover, the possibility of carrying out its
inference in WinBUGS is an important added value, as it makes it
available to a wide community of users for further applications. We
also think that the reparameterization made of the original problem has
also been important, as it has made it possible to incorporate many
previous model selection theory results. These results have given us a
really valuable insight into the prior distributions of noncommon
parameters, which is the main challenge from the Bayesian point of
view, in order to give a reasonable answer to model selection problems.

Our proposal models the observed mortality as Poisson counts. This
distributional assumption has several advantages: (i) it can cope with
zero counts without any problem, (ii) no additional assumption has to
be made about the variability of the observations, since it is
implicitly established and (iii) avoids any Gaussian assumption that
may not be appropriate at all, especially when the observed counts are
lower. On the other hand, the location of change-points is now
considered continuous in time, as in \citet{YuBarretea2007},
 \citet{GhoshBasuea2009}. In our opinion this is a more realistic assumption,
as if a rupture point really has existed, it could have occurred at any
moment in the period under study, regardless of whether we have
observed counts aggregated for subintervals of that same period.

A secondary question is to know how the Bayesian approach compares with
the existing methods in the frequentist arena. We do so through an
intensive simulation study (details provided as supplementary material
[Marti\-nez-Beneito, Garcia-Donato and Salmer{\'{o}}n (\citeyear{Martinez-BeneitoGarcia-Donatoea2011b})]).
They are useful to
know under what circumstances which method is expected to choose, on
average and over replicated data sets, the ``correct'' model. What we
found is that
the Bayesian proposals are more sensitive (compared with the \textit
{Permutation} approach), although less specific in the detection of
joinpoints. This seems to be an expected consequence of the known
conservative behavior of the \textit{Permutation} method and the
probabilistic essence of Bayesian approaches. Therefore, in a context
where data may not be very informative, the use of Bayesian methodology
should really be encouraged. As a consequence, those methods shown in
this paper could be of interest for those disease registries of a
moderate size, not as big as those usually used by the National Cancer
Institute of the United States.

As just outlined, statistical power is an issue of concern in Joinpoint
studies. This concern is even greater when covariates are considered
from the frequentist approach, as in those cases the original data set
is usually split for independent analyses (one for every value of the
covariate) and, as a consequence, the statistical power of every one of
those subanalyses is decreased. The new proposals introduced in this
paper are model-based and, hence, the new covariates could also be
included in our model as main effects or as an interaction with the
terms already considered in the model. Indeed, from the frequentist
point of view some progress has been made toward this kind of modeling
in the case of having at most one joinpoint; see, for example, \citet{PollanPastor-Barriusoea}. In that case we would not be forced to
make independent data analyses for every value of the covariate and
that way we would retain the statistical power of the original analysis
that did not consider any covariate. Also, this kind of analysis would
be the straightforward way to analyze age standardized rates from the
methodology that we have just outlined. This possibility is very
attractive and will be one of the main lines of development following
this study.

\begin{supplement}
\stitle{Supplement Document}
\slink[doi]{10.1214/11-AOAS471SUPP} 
\slink[url]{http://lib.stat.cmu.edu/aoas/471/supplement.pdf}
\sdatatype{.pdf}
\sdescription{A supplemental document for this paper has been written
containing further details about: Performance of the proposed methods
on simulated data sets, Calculus of the basis functions allowing the
fitted trends to describe joinpoints and some remarks about Bayes
factors and their computation in our specific setting.
This document can be found at \citet{Martinez-BeneitoGarcia-Donatoea2011b}.}
\end{supplement}


\printaddresses

\end{document}